\documentclass[prc,preprint,showpacs,floatfix]{revtex4}
\def\beq{\begin{equation}} 
\def\eeq{\end{equation}} 
\begin{document}

\title{ Isovector pairing in a formalism of quartets for N=Z nuclei}

\author{M. Sambataro$^a$ and N. Sandulescu$^b$}
\affiliation{$^a$ Istituto Nazionale di Fisica Nucleare - Sezione di Catania,
Via S. Sofia 64, I-95123 Catania, Italy \\
$^b$ National Institute of Physics and Nuclear Engineering, P.O. Box MG-6, 
Magurele, Bucharest, Romania}

\begin{abstract}
We describe the ground state of the isovector pairing Hamiltonian in self-conjugate nuclei by a product of 
collective quartets of different structure built from two neutrons and two protons 
coupled to total isospin T=0. The structure of the collective quartets is 
determined by an iterative variational procedure based on a sequence of diagonalizations
of the pairing Hamiltonian in spaces of reduced size. The accuracy of the quartet
model is tested for N=Z nuclei carrying valence nucleons outside the
$^{16}$O, $^{40}$Ca, and $^{100}$Sn cores. The comparison with the exact solutions
of the pairing Hamiltonian, obtained by shell model diagonalization,  shows that the quartet 
model is able to describe the isovector pairing energy with very high precision. 
The predictions of the quartet model are also compared to those of the simpler quartet condensation model in which all the collective quartets are
assumed to be identical.


\end{abstract}

\maketitle

In self-conjugate nuclei the isovector proton-neutron  pairing is  
expected to coexist in equal amount with the neutron-neutron and 
proton-proton  pairing, as a consequence of isospin invariance of 
nuclear forces \cite{lane}. The most common approach employed to treat the isovector pairing
in these nuclei is the generalized BCS model \cite{goswami,camiz,ginocchio}. In this approach, however,
the proton-neutron  pairs do not  coexist with the like-particle pairs.
 More precisely, in BCS one usually gets two degenerate solutions for 
the ground state of nuclei with N=Z, namely a solution with only 
proton-neutron  pairs and a solution with only like-particle 
pairs \cite{ginocchio,sandulescu_errea}. As it can be seen from the
comparison with exactly solvable SO(5) pairing models, 
the two BCS solutions do not mix with each other because of the  
isospin symmetry breaking\cite{engel,dobes}. On the other hand, 
restoring the isospin symmetry in
BCS-like models is not enough to obtain a proper description 
of the isovector pairing correlations even if, in addition, the particle number
conservation is also restored \cite{chen}. This fact indicates
the need for a more general isospin conserving formalism which goes beyond the
BCS based approximations. One of such formalisms, explored in 
Ref. \cite{chen}, is the Generator Coordinate Method (GCM) applied on 
a projected-BCS state. An alternative, proposed 
recently in Refs.\cite{qcm1,qcm2}, consists in describing the isovector 
pairing correlations in terms of a condensate of identical 
alpha-like quartets built by two neutrons and two protons coupled to a
total isospin T=0. The aim of this study is to propose a more 
general quartet model approach in which the isovector pairing in the 
ground state of N=Z nuclei is described not as a condensate of identical 
quartets but as a product of quartets having all a different 
structure. It will be shown that by this extension the quartet model gives 
results which almost reproduce the exact solutions of realistic isovector 
pairing Hamiltonians. 

The isovector pairing correlations are described by the  Hamiltonian
\beq
\hat{H}=\sum_i \epsilon_i (N^\nu_i + N^\pi_i) + 
\sum_{i,j} V_{ij} \sum_{\tau}P^+_{i,\tau} P_{j,\tau}.
\eeq
In the first term the operators  $N^\nu_i$ and  $N^\pi_i$ are, 
respectively, the neutron and proton number operators and $\epsilon_i$
are the single-particle energies. Because in this study we treat only systems
with N=Z and since the Coulomb force is not taken into account, the
single-particle energies are considered to be the same for both protons and neutrons.
The pairing interaction is written in terms of the pair operators
\beq
P^+_{i,\tau}=[a^+_i a^+_{\bar i} ]^{T=1}_{\tau}
\eeq
where  ${\bar i}$ stands for the time conjugate of the state $i$ and $\tau$ denotes the three projections of the isospin T=1 corresponding 
to neutron-neutron ($\tau=1$), proton-proton ($\tau=-1$) and proton-neutron
($\tau=0$) pairs. The Hamiltonian (1) with a state-independent  interaction 
strength, i.e., $V_{ij}=g$, can be solved analytically \cite{richardson,pan}.

To describe the ground state of the Hamiltonian (1) for systems with 
N=Z we shall use as building blocks not Cooper pairs, as done in BCS-type models, 
but collective quartets 
formed by two neutrons and two protons. Thus, we
first introduce a set of  non-collective quartets composed by two isovector 
pairs coupled to T=0 
\beq
A^+_{ij} = [P^+_i P^+_j]^{T=0} = \frac{1}{\sqrt 3}
(P^+_{i,1} P^+_{j,-1}+P^+_{i,-1} P^+_{j,1}
           -P^+_{i,0} P^+_{j,0}).
\eeq
With these non-collective quartets we then construct the  collective quartets
\beq
Q^+_{\nu} = \sum_{i,j} q^{(\nu )}_{ij} A^+_{ij}.
\eeq
Finally, with these collective quartets, we define the state
\begin{equation}
|\Psi_{gs}\rangle =\prod^{N}_{\nu =1}Q^\dag_\nu |0\rangle .
\label{200}
\end{equation}
This quartet state provides our ansatz 
for the ground state of the isovector pairing Hamiltonian (1) in the case
of even-even proton-neutron systems with N=Z.
It is worth mentioning that the present quartet model is different from the roton model of  Arima and Gillet \cite{arima} whose quartets, as well as the proton and neutron pairs explicitly forming them, are allowed to couple to any angular momentum. Due to such a feature, the roton model appears to be more suited to describe the quarteting in heavy deformed nuclei where protons and neutrons occupy different valence shells and the proton-neutron pairing in not expected to play any important role.

The quartet state (5) depends on the mixing amplitudes  which
define the collective quartets (4).  In order to search for the most 
appropriate $q^{(\nu )}_{ij}$'s we make use of an iterative variational 
procedure. This procedure is basically identical to that we have recently used 
in the case of like-particle quartets \cite{sasa} and draws inspiration 
from an analogous technique previously developed for a treatment of pairing 
correlations in terms of a set of independent pairs \cite{samba2}. 
The procedure consists of a sequence of diagonalizations of the Hamiltonian in 
spaces of a rather limited size. More in detail, let us suppose that, at a given stage 
of the iterative process, one knows the state (\ref{200}) and let us construct the space 
\begin{equation}
F^{(\rho)}_N=\Biggl\{ [P^\dag_iP^\dag_{j}]^{T=0} 
\prod^{N-1}_{\nu=1(\nu\neq \rho)}Q^\dag_\nu |0\rangle 
\Biggr\}_{1\leq i\leq j\leq\Omega}
\label{5}
\end{equation}
($\Omega$ being the number of single-particle states).
The states of $F^{(\rho)}_N$ are generated by acting with the operators 
$[P^\dag_iP^\dag_j]^{T=0}$ on the product of all the quartets
$Q^\dag_\nu$ but the $\rho$-th one. The dimension of each space (\ref{5})   
is therefore at most $\Omega (\Omega +1)/2$ and one can form $N$ such spaces. 
By diagonalizing the Hamiltonian in $F^{(\rho)}_N$ and searching for the 
lowest eigenstate, one constructs the state
\begin{equation}
|\Psi_{gs}^{(new)}\rangle =Q^{\dag (new)}_\rho\prod^{N-1}_{\nu=1(\nu\neq \rho)}Q^\dag_\nu |0\rangle 
\label{6}.
\end{equation}
This differs from $|\Psi_{gs}\rangle$ only for the new quartet $Q^{\dag (new)}_\rho$ and its
energy is by construction lower than (or, at worst, equal to) that of 
$|\Psi_{gs}\rangle$. As a result of this operation, the 
quartet $Q^{\dag (new)}_\rho$ has updated $Q^{\dag}_\rho$ while all the other quartets 
have remained unchanged. At the same time the energy of  $|\Psi_{gs}\rangle$ has been 
driven towards its minimum. Performing a series of diagonalizations 
of $H$ in $F^{(\rho)}_N$ for all possible $\rho$ values $(1\leq\rho\leq N)$ exhausts 
what we define an iterative cycle. At the end of a cycle all the 
quartets $Q^\dag_\nu$ have been updated and a new cycle can then start. 
The sequence of iterative cycles goes on until the difference between the ground 
state energy at the end of two successive cycles becomes vanishingly small.
In practice, in order to describe a system with $N$ quartets, one proceeds 
step-by-step starting from the case of just one quartet. 
The diagonalization of $H$ in the space $F^{(1)}_1$, which simply consists of the 
states $[P^\dag_iP^\dag_{j}]^{T=0}|0\rangle$, generates $Q^\dag_1$. For $N=2$, 
the diagonalization in $F^{(2)}_2$, with $Q^\dag_1$ taken from the previous calculation, 
generates the first approximation of the quartet $Q^\dag_2$. Iterating these diagonalizations 
in $F^{(1)}_2$ and $F^{(2)}_2$ gives rise to the final 
quartets $Q^\dag_1$ and $Q^\dag_2$ for $N=2$. Similarly, for $N=3$, 
the diagonalization in $F^{(3)}_3$, with $Q^\dag_1$ and $Q^\dag_2$ taken from the 
previous calculation, generates the first approximation of 
the quartet $Q^\dag_3$. And so on and so forth.

In the quartet state (5) the quartets are different from one another. A simpler approach,
to which the quartet model will be compared below, simply assumes that all quartets 
have the same structure \cite{qcm1}. In this approach, called  quartet condensation
model, the ground state is therefore approximated by the quartet condensate
\begin{equation}
| \Psi \rangle =(A^+)^{N} |0 \rangle ,
\end{equation}
where $A^+$ is the collective quartet defined by 
\beq
A^+ = \sum_{i,j} x_{ij} A^+_{ij}.
\eeq
To make possible the connection to the Cooper pairs used in BCS-type models, 
in Ref. \cite{qcm1} it was supposed that the mixing amplitudes 
of the collective quartet (9) are separable, i.e., $x_{i,j}=x_i x_j$. 
With this approximation the collective quartet can be written as 
\beq
A^+= 2 \Gamma^+_1 \Gamma^+_{-1} - (\Gamma^+_0)^2,
\eeq
where $\Gamma^+_{t}= \sum_i x_i P^+_{i,t}$, for t={0,1,-1},  denote the 
collective pair operators for the proton-neutron, neutron-neutron and 
proton-proton pairs, respectively. Due to the isospin invariance, all the collective pairs 
have the same mixing amplitudes $x_i$. They are determined by minimizing the
average of the Hamiltonian (1) with the normalization constraint. The details of the 
calculations can be found in Ref. \cite{qcm1}.

\begin{table}[hbt]
\caption{ Correlation energies for spherical single-particle states and pairing
forces extracted from standard shell model interactions.
The results are shown for the exact  diagonalizations, the quartet model (QM)
and the quartet condensation model (QCM) of Ref.\cite{qcm1}. 
In brackets we give the errors relative to the exact results.}
\begin{center}
\begin{tabular}{|c|c|c|c|}
\hline
\hline
   &    Exact & QM & QCM  \\
\hline
\hline
$^{20}$Ne  &  9.174   & 9.174 (-)       & 9.170  (0.04\%)   \\
$^{24}$Mg  &  14.461  & 14.458 (0.02\%)   & 14.436 (0.17\%)   \\
$^{28}$Si  &  15.787  & 15.780 (0.04\%)  & 15.728 (0.37\%)   \\
$^{32}$S   &  15.844 & 15.844 (-)  & 15.795 (0.31\%)  \\
\hline
$^{44}$Ti  &  5.965   & 5.965 (-)       & 5.964 (0.02\%)   \\
$^{48}$Cr &   9.579   & 9.573 (0.06\%)   & 9.569 (0.10\%)    \\
$^{52}$Fe &  10.750   & 10.725 (0.23\%)  & 10.710 (0.37\%)   \\
\hline
$^{104}$Te &  3.832    & 3.832 (-)           & 3.829 (0.08\%)    \\
$^{108}$Xe &  6.752   & 6.752 (-)          & 6.696 (0.83\%)   \\
$^{112}$Ba &  8.680    & 8.678 (0.02\%)          & 8.593 (1.00\%)   \\

\hline
\hline
\end{tabular}
\end{center}
\end{table}

To test the accuracy of the quartet model we have performed calculations 
for three sets of N=Z nuclei with valence nucleons 
outside the cores $^{16}$O, $^{40}$Ca, and $^{100}$Sn. 
Following Ref.\cite{qcm1}, we have performed calculations with two different 
inputs for the isovector pairing Hamiltonian (1) and choosing only
the N=Z systems  for which the Hamiltonian can be diagonalized exactly. 
We have first considered the case of spherically symmetric single-particle
states and isovector pairing forces extracted from the (T = 1, J = 0) part 
of standard shell model interactions. Namely, for the three sets of nuclei 
mentioned above we  have extracted the isovector pairing force from the  
universal sd-shell interaction USDB \cite{usd},  the monopole-modified 
Kuo-Brown interaction KB3G \cite{kb3g} and, respectively, the 
effective G-matrix interaction of Ref. \cite{gmatrix}.
Details about the single-particle energies employed in the calculations
are given in Ref. \cite{qcm1}.
The results for the pairing correlations energies, defined as the difference
between the ground state energies obtained without and with the
pairing force, are given in Table I. The correlations energies predicted by the
quartet model (QM)  are compared to the exact results and the results of 
quartet condensation model (QCM). The QCM results and the
exact results \cite{error} are extracted from Ref. \cite{qcm1}.
We notice that for the systems with one quartet outside the closed core 
the quartet state (5) is by construction exact. This is not the case for the 
quartet condensate state (8) because of the factorization approximation 
$x_{ij} = x_i x_j$.  For systems with more than one quartet outside the core 
the quartet state (5) is not anymore exact. However, as seen in Table I,  
the errors relative to the exact solution are very small. We can also observe that 
QM gives smaller errors than QCM, reflecting the gain in correlation energy obtained 
in QM by allowing the quartets to be different.

\begin{table}[hbt]
\caption{ Correlation energies calculated for axially deformed single-particle states
and a state-independent isovector pairing force.
The results are shown for the exact  diagonalizations, the quartet model (QM)
and the quartet condensation model (QCM) of Ref.\cite{qcm1}. 
In brackets we give the errors relative to the exact results.}
\begin{center}
\begin{tabular}{|c|c|c|c|}
\hline
\hline
   &    Exact & QM & QCM  \\
\hline
\hline
$^{20}$Ne  &  6.5505  & 6.5505      & 6.539  (0.18\%)   \\
$^{24}$Mg  &  8.4227 & 8.4227      & 8.388 (0.41\%)   \\
$^{28}$Si  &  9.6610 & 9.6610      & 9.634 (0.28\%)    \\
$^{32}$S   &  10.2629 & 10.2629     & 10.251 (0.12\%)  \\
\hline
$^{44}$Ti &  3.1466 & 3.1466      & 3.142 (0.15\%)    \\
$^{48}$Cr &  4.2484 & 4.2484      & 4.227 (0.50\%)     \\
$^{52}$Fe &  5.4532 & 5.4531      & 5.426 (0.50\%)      \\
\hline
$^{104}$Te &  1.0837 & 1.0837      & 1.082 (0.16\%)    \\
$^{108}$Xe &  1.8696 & 1.8696      & 1.863 (0.35\%)     \\
$^{112}$Ba &  2.7035 & 2.7034      & 2.688 (0.57\%)  \\

\hline
\hline
\end{tabular}
\end{center}
\end{table}

We have also tested the accuracy of the quartet model for an isovector pairing 
interaction acting on the single-particle spectrum corresponding to an 
axially deformed mean field. As in Ref. \cite{qcm1}, the single-particle
energies have been extracted from axially deformed Skyrme-HF calculations
performed with the force SLy4 \cite{sly4} and neglecting the Coulomb interaction. 
For the  isovector pairing force we have taken a state-independent interaction
with strength $g$= 24/A. As in the case of spherical symmetry, the pairing has been 
applied to the nucleons outside the cores $^{16}$O, $^{40}$Ca, and $^{100}$Sn. 
In the calculations we have considered, respectively, the  lowest seven, nine, and ten HF single-particle states above the cores just mentioned. The number of these states has been chosen such as to keep the total degeneracy of the model space approximately the same as in the case of spherical calculations.
The results are presented in Table II. In this case the correlation energies predicted by the quartet model are basically
exact. The same high precision has been observed for the occupation probabilities. 

In order to test its accuracy, the quartet model has been applied here only to N=Z systems for which the Hamiltonian (1) can be diagonalized exactly by shell model techniques. However, we remark that, being based on diagonalizations of matrices of reduced size and also involving a model space much smaller than the full shell model space, the quartet model lends itself to the treatment of N=Z nuclei with more valence nucleons and/or larger shells than in standard shell model calculations. In addition, the quartet model provides a deeper insight into the structure of the ground state wave function.

In conclusion, in this study we have proposed a quartet model for the ground 
state of the isovector pairing Hamiltonian in self-conjugate nuclei. This model assumes that the ground
state of even-even systems with N=Z is a product of collective, distinct T=0 quartets built 
by two neutrons and two protons. The collective quartets are determined by an iterative variational procedure.
The calculations carried out for various isovector pairing Hamiltonians have shown that 
the quartet model reproduces with very high precision the ground state correlation energies of these systems. 
This model therefore proposes itself as a very appropriate tool for the treatment of the isovector pairing in 
mean field type models. We also like to emphasize that the quartet formalism discussed in this work can be 
extended in a straightforward way to the treatment of more complex Hamiltonians such as, for instance, 
the isovector plus isoscalar pairing Hamiltonian. Work is in progress in this direction.

\vskip 0.3cm

{\it Acknowledgments} 
The authors are very grateful to Calvin Johnson for providing revised shell model results.
This work was supported by the Romanian Ministry of Education and Research
through the grant Idei nr 57.


\begin{thebibliography}{10}
\bibitem{error}
In Refs. \cite{qcm1, qcm2},
the exact results for the pf-shell nuclei (spherical case) are not correct  
due to an input error in the shell model calculations.
\end{thebibliography}
\end{document}